\documentclass[aps,prl,showpacs,twocolumn,amsmath,amssymb]{revtex4}
\usepackage{graphicx,bbm}

\newcommand{\bs}[1]{{\boldsymbol #1 }}
\newcommand{\ave}[1]{\langle #1 \rangle}

\usepackage{graphicx}
\usepackage{dcolumn}
\usepackage{bm}
\usepackage[dvips]{color} 

\begin{document}

\preprint{}

\title{
Nonequlibrium particle and energy currents in quantum chains connected to mesoscopic Fermi reservoirs
}

\author{Shigeru Ajisaka} \email{g00k0056@suou.waseda.jp}
\affiliation{Departamento de F\'isica, Facultad de Ciencias F\'isicas y
Matem\'aticas, Universidad de Chile, Casilla 487-3, Santiago Chile}
\author{Felipe Barra}
\affiliation{Departamento de F\'isica, Facultad de Ciencias F\'isicas y
Matem\'aticas, Universidad de Chile, Casilla 487-3, Santiago Chile}
\author{Carlos Mej\'ia-Monasterio}
\affiliation{Laboratory of Physical Properties, Technical University of Madrid, Av. Complutense s/n 28040, Madrid, Spain}
\author{Toma\v{z} Prosen}%
\affiliation{Faculty of Mathematics and Physics, University of Ljubljana, Jadranska 19, SI-1000 Ljubljana, Slovenia}

\date{\today}
\begin{abstract}
  We propose a model of nonequilibrium  quantum transport of particles
  and energy in a system connected to mesoscopic Fermi reservoirs (meso-reservoir). The
  meso-reservoirs are in turn thermalized to prescribed temperatures and
  chemical potentials by a simple dissipative mechanism described by the
  Lindblad equation.  As an example, we study transport in monoatomic and diatomic chains of
  non-interacting spinless fermions. 
  We show numerically the breakdown
  of the Onsager reciprocity relation due to the dissipative terms of the model.
\end{abstract}


\maketitle

{\em Introduction.-} Nonequilibrium systems abound in nature and still
their theoretical description poses numerous challenges to theory.
Nonequilibrium steady states (NESS) describe the state of a system
maintained out of equilibrium by external forces such as {\it e.g.},
gradients of temperature $T$ and chemical potential $\mu$, and are
characterized by the emergence of steady flows. One might expect that
a finite system connected to two infinitely extended reservoirs
imposing external gradients, will reach a NESS after a sufficiently
long time~\cite{Rue00,Rue01}. Nonetheless, this is not always the case
and the conditions to reach a NESS are not well understood in general.
Although the problem of existence and approach to NESS has been
discussed since the early days of statistical mechanics, rigorous
results are limited to a few examples \cite{RLL,KMP,Schutz},
and to near equilibrium regimes. The extension to regimes beyond
linear response faces the difficulty arising from infiniteness of the
reservoirs~\cite{Zubarev95,ST06}.

In quantum mechanics the construction of NESS requires to consider
open quantum systems, rendering the problem extraordinarily more
difficult. The common setup is to consider the infinite time limit of
the density matrix of the finite system $S$ coupled to two infinite
reservoirs which are in thermal equilibrium at different temperatures
and chemical potentials, starting from initial separable state
$\rho_L\otimes \rho_S\otimes \rho_R$.  One can then study the
properties of the density operator of the total (infinite) system
\cite{Rue00,Rue01}, or the reduced density operator for the (finite)
system~$S$ \cite{Breuer02}, obtained by tracing out the reservoirs'
degrees of freedom.  The time evolution of the density operator is
naturally determined by the Von Neumann equation.  However, dealing
with infinite degrees of freedom is in most cases difficult. A second
approach based on the master equation of the reduced density operator
is more accessible, albeit at the price of several approximations such
as {\it e.g.}, Born-Markov (see e.g.  \cite{Wichterich07}). 
%

\begin{figure}
\includegraphics[width=80mm]{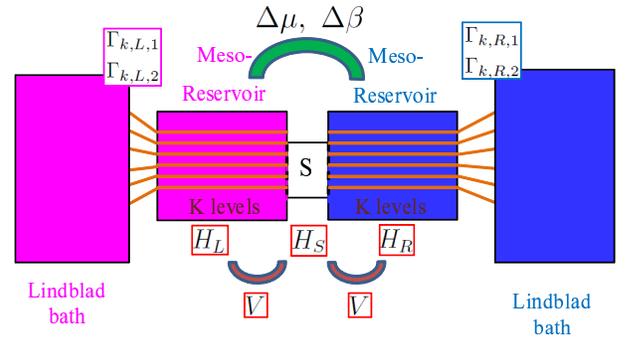}
\vskip -4mm
\caption{\label{figure1} (Color online) The schematic view of the
  two-level reservoir open system model. }
\end{figure}

In this Letter, we propose a conceptually different approach for the
study of NESS that comprises meso-reservoirs with a finite
number of degrees of freedom, which in turn are thermalized by
Markovian macroscopic reservoirs using a simple Lindblad equation.
This setup yields a computationally efficient model in which the NESS
of the system and of the meso-reservoirs are mathematically accessible, thus
allowing to study characteristic features depending on the number of
degrees of freedom of the meso-reservoirs, such as relaxation times of the
system and of the meso-reservoirs, or correlations among them. It is
interesting remarking that finite meso-reservoirs in principle allow the
creation of entanglement among the different degrees of freedom of the
system solely through its coupling with the meso-reservoir \cite{Braun},
thus making our setup much richer.


{\em The model.-} We consider a one-dimensional quantum chain of
spinless fermions coupled at its boundaries to meso-reservoirs comprising a
finite number of spinless fermions with wave number $k$
($k\in\{1,\ldots,K\}$). The Hamiltonian of the system can be written
as $H = H_S+H_L+H_R+V$, where
\begin{eqnarray} \label{def:H}
H_S &=& -\sum_{j=1}^{n-1} \Big(t_j c_j^\dag c_{j+1} + ({\rm h.c.}) \Big)
+\sum_{j=1}^{n} U c_j^\dag c_j \ ,
\\
H_{\alpha} &=& \sum_{k=1}^{K} 
\epsilon_k a_{k\alpha}^\dag a_{k\alpha}
,\ \ 
\epsilon_k \equiv \theta_F (k-k_0)
,\ \ \alpha\!=\!\{L, R\} 
\nonumber
\\
V &=& \sum_{k=1}^{K} v\left( a_{kL}^\dag c_1 + a_{kR}^\dag c_n\right) + ({\rm h.c.}) \ ,
\nonumber
\end{eqnarray}
and  where $\{t_j\}$  are the  nearest neighbor  hoppings, $U$  is the
onsite  potential, $v$  is the  coupling  between the  system and  the
meso-reservoirs,   and  $c_j,c_j^\dagger$   is   the  annihilation/creation
operator    for    the    spinless    fermions   of the chain,    while
$a_{k,\alpha},a^\dagger_{k,\alpha}$  are that  of the  left  and right
meso-reservoirs.

The  key idea  of our  model is  to enforce  the  finite meso-reservoirs to     equilibrium using simple Lindblad
dissipators~\cite{Koss76,Lindblad76,Kosov11}.  If  the   couplings   to  the
Lindblad dissipators are  small, then we can interpret  these terms as
coming  from   tracing  out  infinitely   extended  (super-)reservoirs
(schematically  depicted  in  Fig.~\ref{figure1}). Thus,  the  density
matrix of the total setup  evolves according to the many-body Lindblad
equation
\begin{eqnarray}
\frac{{\rm d}}{{\rm d}t}\rho &=&
-i[H,\rho] \label{eq:lindbladmodel}
\\
&&+\sum_{k, \alpha, m}
\left(
2L_{k, \alpha, m} \rho L_{k, \alpha, m}^\dag
-\{L_{k, \alpha, m}^\dag L_{k, \alpha, m},\rho\}
\right)
\nonumber \\
L_{k,\alpha,1} &=& \sqrt{ \Gamma_{k,\alpha,1} } a_{k\alpha}
,\ \
L_{k,\alpha,2}=\sqrt{ \Gamma_{k,\alpha,2} } a_{k\alpha}^\dag
\nonumber\\
\Gamma_{k,\alpha,1}&=&\gamma ( 1 - F_{\alpha}(\epsilon_k) )
,\ \ 
\Gamma_{k,\alpha,2}=\gamma F_{\alpha} (\epsilon_k)
\nonumber\ ,
\end{eqnarray}
where $L_{k,\alpha,1}$ and $L_{k,\alpha,2}$ are operators  representing the coupling to the
super-reservoirs,                  $F_\alpha(\epsilon)=(e^{\beta_\alpha
  (\epsilon-\mu_\alpha)}+1)^{-1}$   are   Fermi  distributions,   with
inverse   temperatures    $\beta_\alpha$   and   chemical   potentials
$\mu_\alpha$, and  $[\cdot , \cdot]$  and $\{ \cdot ,  \cdot\}$ denote
the       commutator        and       anti-commutator,
respectively. Generalization to  spinfull fermions (e.g. electrons) is
straightforward.
The parameter $\gamma$ determines the strength of the coupling to the
super-reservoirs and, as we shall show later, needs to be fine-tuned
in order to ensure the applicability of the model.  We note that the parameter $\gamma$ determining the rate of relaxation of
the meso-reservoirs to equilibrium, in general can depend on
temperature, chemical potential, and wave-number $k$ of the meso-reservoir
modes. Here we consider the simplest model possible and take $\gamma$
as constant.  We stress that our model does not rely on the usual
weak-coupling assumption needed for the physical derivation of the
Lindblad master equation \cite{Breuer02}, thus $\gamma$ does not need
to be a small parameter.

{\em  Results.-} We  have  studied monoatomic  ($t_j=t$) and diatomic
($t_{2j-1}=t_A,\ t_{2j}=t_B$)  chains. Unless specified  differently, we
have  set   $\gamma=0.1,\  v=0.03,\  \epsilon_1=-20,\  \epsilon_K=20,\
K=200$, $t=3$, and $\mu_L=-\mu_R=\mu$.

\begin{figure}
 \includegraphics[width=88mm]{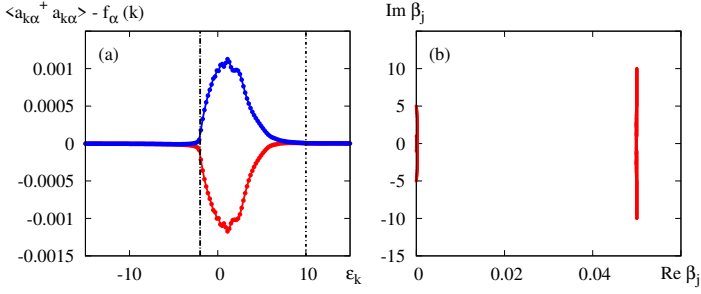}
\vskip -4mm
\caption{\label{figure2} (Color online)
 Panel (a) shows (mode)
  $k$-dependence of the deviation from the Fermi function
  $\ave{a^\dag_{k\alpha} a_{k\alpha}}-F_\alpha(\epsilon_k)$, 
(red/blue curve is for the left/right meso-reservoir,  
and vertical lines indicate the energy band, see main text for parameter values), and Panel (b) shows the rapidity
  spectrum for the monoatomic chain indicating the
  separation of time scales. }
\end{figure}

When the system and the meso-reservoirs are decoupled ($v=0$),
each non-interacting mode of the meso-reservoir is thermalized separately
with the prescribed Fermi-Dirac occupation number~\cite{Prosen08}.
For small $v$, $v \ll \gamma$, we thus expect the distribution of occupations in the
meso-reservoirs to be close to Fermi-Dirac, as shown in
Fig.~\ref{figure2}(a).  A remarkable feature of our model is that we
can monitor and control the difference of occupation distributions
$\langle a_{k\alpha}^\dag a_{k\alpha}\rangle-F_\alpha(\epsilon_k)$ by
changing the coupling parameters $v$ or
$\gamma$.
Here and bellow $\ave{\cdot}$
represents an expectation value with respect to the NESS.

Looking at the occupation density inside the system we observe that
$\ave{c_j^\dag c_j}$ is almost a constant, i.e. independent of $j$,
except at the edge of the chain, indicating that ballistic transport
is expected.


We now discuss the relaxation times of the system.  As shown in
\cite{Prosen10}, the spectrum of the evolution superoperator is given
in terms of the eigenvalues $\beta_j$ (so-called rapidities) of the matrix $X$:
\begin{eqnarray*}
\bs{X} &=& -\frac{\rm i}{2}\bs{H}\otimes \sigma_y + 
	\frac{\gamma}{2}
	\begin{pmatrix}
	\bs{E}_{K} & \bs{0}_{K\times n} &\bs{0}_{K\times K}\\
	\bs{0}_{n\times K} & \bs{0}_{n\times n} &\bs{0}_{n\times K}\\
	\bs{0}_{K\times K} & \bs{0}_{K\times n} &\bs{E}_{K}
	\end{pmatrix}
	\otimes\bs{E}_2
\end{eqnarray*}
where $\bs{0}_{i\times j}$ and $E_j$ denote $i\times j$ zero matrix
and $j\times j$ unit matrix, $\sigma_y$ is the Pauli matrix, and
$\bs{H}$ is a matrix which defines the quadratic form of the
Hamiltonian, as $H= \bs{d}^T \bs{H} \bs{d}$ in terms of fermionic
operators $\bs{d}^T \equiv \{a_{1L},\cdots
a_{KL},c_1,\cdots,c_n,a_{1R},\cdots a_{KR} \}$.  Fig.~\ref{figure2}(b)
shows a typical spectrum for the monoatomic chain.  Interestingly, there
is a clear separation of the relaxation times into slow and fast
normal modes.  The number of slow modes $n_s$, with eigenvectors
localized in the system part, is $n_s \approx 2n$, and the number of
fast modes $n_f\approx 4 K$, with eigenvectors localized in the
meso-reservoirs. 

We have not observed rapidities with zero real part, and thus
according to Ref.~\cite{Prosen08,Prosen10}, there exists a unique
steady state for the range of parameters used in this Letter.  Next, 
we study the NESS averages $\ave{\cdot}$ of
observables. The quadratic observables are given in terms of the
solution of the Lyapunov equation \cite{Prosen10}:
\begin{eqnarray*}
&&\ave{ w_j w_k}
=\delta_{j,k} -4iZ_{j,k}
\\
&&
w_{2j-1}\equiv c_j+c_j^\dag,\ \ w_{2j}\equiv i(c_j-c_j^\dag)
\\
&&
\bs{X}^T \bs{Z}+\bs{Z}\bs{X} \equiv  \bs{M}_i
\\
&&\bs{M}_i \equiv -\frac{i}{2} {\rm diag}\{\Gamma^-_{1L},\cdots,\Gamma^-_{KL},
\bs{0}_{1\times n},\ \Gamma^-_{1R},\cdots,\Gamma^-_{KR}\}
\otimes \bs{\sigma_y}
\\
&&\Gamma^-_{k\alpha} \equiv \Gamma_{k, \alpha, 2} - \Gamma_{k, \alpha, 1}
=\gamma_{k\alpha}\{ 2F_\alpha (\epsilon_k)-1 \}\ \ ,
\end{eqnarray*}
whereas Wick's theorem can be used to obtain expectations of
higher-order observables.
Let us now discuss the deviation of occupation numbers from the Fermi distribution in the meso-reservoirs.  
Lindblad equation of motion results in the following identity, 
$2\gamma \left\{ \ave{a^\dag_k a_k}-F_\alpha (\epsilon_k) \right\}
=\ave{J^\alpha_k}$, where $J^\alpha_k$ is the $k$-th level contribution to the current from the $\alpha$-meso-reservoir to the system, $J^\alpha_k=iv(a_k^\dag c_\alpha-c_\alpha^\dag a_k)$ 
($c_L=c_1,\ c_R\equiv c_n)$. 
This relation can be used to determine the Lindblad dissipator from physical observables such as currents and occupation numbers.
It is interesting to remark that this relation can be interpreted as a type of Landauer formula, i.e., 
$\ave{J^P_{tot}}=\sum_k  2\gamma \{ \ave{a^\dag_{kL} a_{kL}}-F_L(\epsilon_k)\}$, where $J^P_{tot}$ is the particle current from the left reservoir to the system, and shows explicitly how the nonequilibrium situation modifies the Fermi distributions. A similar expression holds for the
current from the right reservoir to the system.  
It follows that although the distribution functions $\langle
a_{k\alpha}^\dag a_{k\alpha}\rangle$ are slightly modified by the presence of the
system and the other reservoir, the integrated difference from the
Fermi distribution summed over both meso-reservoirs satisfies the
charge conservation, i.e., $\sum_{k,\alpha} \left\{\langle
  a_{k\alpha}^\dag a_{k\alpha}\rangle-F_\alpha(\epsilon_k)\right\}=0$
  \footnote{Nevertheless, we find that even a slightly more general relation is satisfied in NESS of the system with arbitrary $K$-level meso-reservoirs,
  namely having two possibly different rates of coupling to Lindblad super-reservoirs, $\gamma_L$ and $\gamma_R$, we have
  $\sum_{k,\alpha} \gamma_\alpha \left\{\langle a_{k\alpha}^\dag a_{k\alpha}\rangle-F_\alpha(\epsilon_k)\right\}=0$. For $\gamma_L\neq \gamma_R$, particle conservation is not satisfied.}.

We now turn our attention to the particle and energy currents in the system.  The particle current
is defined through the conservation law of number of particles:
\begin{eqnarray*}
\frac{{\rm d} c_j^\dag c_j}{{\rm d}t} &=& J^P_{j-1}-J^P_{j},\ (2\le j \le n-1)
\\
J^P_{j} &\equiv& it_{j}(c^\dag_{j} c_{j+1} - c^\dag_{j+1}c_{j})
,\ (1 \le j\le n-1)
\end{eqnarray*}
and the energy current is defined through the conservation law of
local energy:
\begin{eqnarray*}
&&\frac{{\rm d} H_j }{{\rm d}t}=J^E_{j-1}-J^E_{j},\ (2\le j\le n-1)
\\
&&J^E_{j-1}=i[H_{j-1},H_j]
,\ \ (2\le j\le n)
\\
&&H_j \equiv t_j c_j^\dag c_{j+1} + t_j c_{j+1}^\dag c_j
+U c_j^\dag c_j
,\ (1\le j\le n)
\end{eqnarray*}
where, by definition, $c_{n+1}\equiv 0$.  We have checked that both
the particle and the energy current converge to a constant by
increasing the system size $n$ for the monoatomic chains.  For the
diatomic chains, currents converge to different constants, depending
on the parity of integer $n$.  Therefore, ballistic transport is
indeed achieved.

\begin{figure}
 \includegraphics[width=80mm]{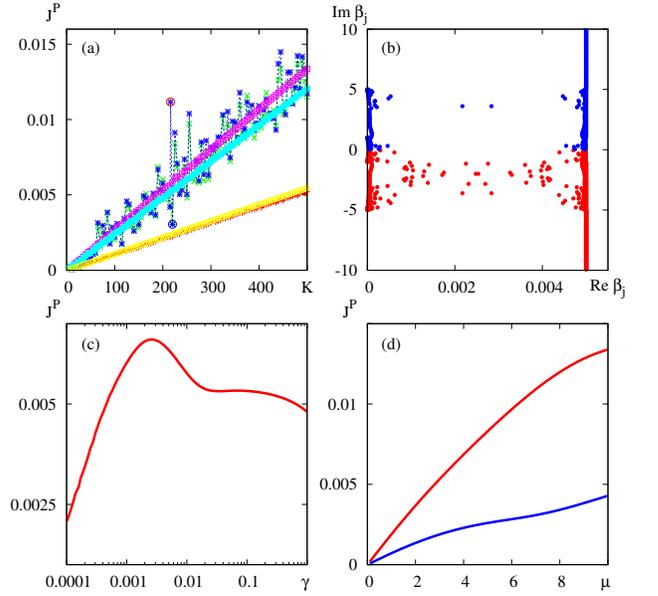}
\vskip -4mm
  \caption{
\label{figure3}
(Color online) Panel (a) shows the $K$ dependence of the particle
current for the monoatomic chains (red: $\gamma=0.0001$, green:
$\gamma=0.001$, blue: $\gamma=0.01$, pink: $\gamma=0.1$, cyan:
$\gamma=1$, yellow: $\gamma=10$), and panel (b) shows the rapidities for 
$\gamma=0.01$ (Using the symmetry with respect to real axis, we plot rapidities in positive/negative imaginary plane for $K=216$/$K=220$). Those two parameters are indicated in figure (a) as red ($K=216$) and blue ($K=220$) circles.
Panel (c) shows the $\gamma$ dependence
of the particle current for the monoatomic chains. 
Panel (d) shows $\mu$ dependence of the particle current. The red line is
for the monoatomic chains ($t=3$), and the blue line is for the
diatomic chains ($t_A=3,\ t_B=6$).}
\end{figure}

Fig.~\ref{figure3}(a) shows the $K$ dependence of the particle current
for the monoatomic chains with $\gamma$ ranging from $0.0001$ to $10$.
The particle current increases linearly in the number of meso-reservoir
modes $K$, with notable fluctuations in $K$ observed for some range of
$\gamma$ (which will be discussed later).  We see that the particle
current is roughly $\gamma$-independent for $\gamma \in [0.001,1]$,
and is somewhat smaller for very small or very large $\gamma$.
The dependence on $\gamma$ of the particle current, shown in
Fig.~\ref{figure3}(c), exhibits a plateau starting at $\gamma \approx v$, and one should take
$\gamma$ in the plateau region $[v, \|H\|]$ to have a physically meaningful model. 
\begin{figure}
 \includegraphics[width=80mm]{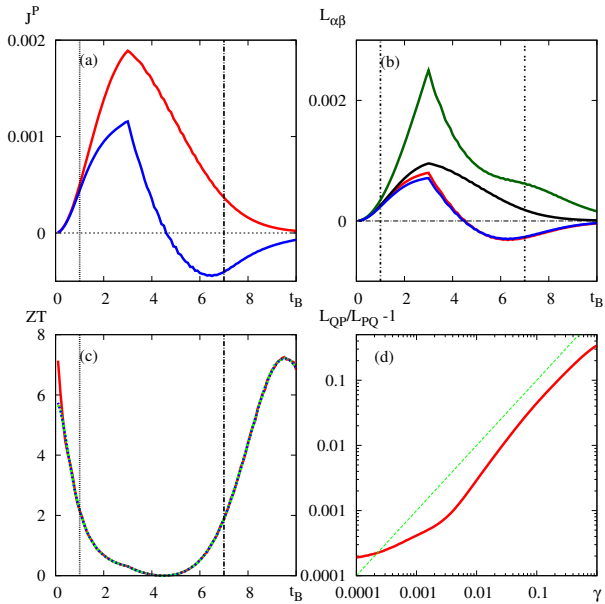}
\vskip -4mm
\caption{\label{figure4} (Color online)
 Panel (a) shows the $t_B$
dependence of the particle current for the chains.  The red
line is for $\beta_L=\beta_R=1$, and $\mu=1$, and the blue line is
for $\beta_L=0.5,\ \beta_R=4$, and $\mu=0$.
Panel (b) shows $t_B$ dependence of the
  Onsager coefficients (color code:  $L_{QP}$ - red,  $L_{PP}$ - black, 
  $L_{QQ}$ - green, $L_{PQ}$ - blue), and (c) of $ZT$ (comparing $K=200$ case (green) to $K=100$ (red) and $K=500$ (blue)), for the diatomic chains
  ($t_A=3$). Chains have exponentially small current outside $t_b\in [1,7]$ indicated in the figure. 
Panel (d) shows the
  $\gamma$ dependence of asymmetry $L_{QP}/L_{PQ}-1$ for the diatomic chain with
  $t_A=3,\ t_B=6$ (dashed line indicates linear growth). Note that asymmetry grows linearly with $\gamma$ in the physical regime $\gamma > v$.}
\end{figure}
Fig.~\ref{figure3}(d) shows the dependence of the particle current on
the bias of the chemical potential ($\mu$). We find, as expected,
initially linear increase in $\mu$ which slows down for larger $\mu$.

The fluctuations in the current as a function of $K$ occur because for
some values of $K$ the coupling between the meso-reservoirs and system is
stronger. Intuitively this happens when an eigenenergy of the
meso-reservoir coincides with one of the system, resulting with some
eigenvector of the matrix $\bs{X}$ being supported both in the system
and the meso-reservoirs, breaking the separation into slow and fast modes.
When the density of states of the meso-reservoir becomes dense enough such
that it is smooth (average level spacing smaller than the width given
by the coupling $\gamma$ to the super-reservoirs), the current becomes
a smooth function of $K$ or equivalently of the density of
states. These ideas are confirmed by the numerical results in
Fig.~\ref{figure3}(a,b). We note however, the current fluctuations are strong only in the `non-physical' regime $\gamma < v$.

Fig.~\ref{figure4}(a) shows the dependence of the particle current
driven by either thermal or chemical gradients, on the hopping
strength ($t_B$) for diatomic chains.  
Where the particle current is negative, the energy band of the chain
is located entirely below the fermi energy of the cooler reservoir,
yielding a particle current flow from the cold to the hot
reservoir. Such crossed transport (and its counterpart of the energy
current driven by the chemical gradient), can be exploited to pump
heat or particles \cite{casati}.

For sufficiently  small thermal  and chemical gradients,  the particle
and   heat   current   defined   as  $J_Q\equiv   J_E-\bar{\mu}   J_P$
($\bar{\mu}=(\mu_L+\mu_R)/2$),   depend  linearly   on   the  external
gradients as \cite{Domenicali54,Callen48,Groot84}
\begin{eqnarray*}
J_Q &=& L_{QQ}\Delta \beta+L_{QP} \Delta (-\beta \mu) \ ,
\\
J_P &=& L_{PQ}\Delta \beta+L_{PP} \Delta (-\beta \mu) \ ,
\end{eqnarray*}
where  $\Delta  \beta\equiv  \beta_R-\beta_L$ and  $\Delta  (\beta\mu)
\equiv \beta_R\mu_R-\beta_L\mu_L$.   The second law  of thermodynamics
imposes  definite-positiveness of the  matrix of  Onsager coefficients
$\bs{L}$, which  implies $L_{QQ}\ge 0$  and $L_{PP}\ge 0$, and  if the
dynamics  is  time-reversible,   the  Onsager's  reciprocity  relation
$L_{PQ}=L_{QP}$  holds.  In  Fig.~\ref{figure4}  we consider diatomic
chains with $t_A=3$  and show the dependence of  various properties of
$\bs{L}$ on the  other hopping parameter $t_B$.  Fig.~\ref{figure4}(b)
shows  the $t_B$ dependence  of all  Onsager coefficients,  whereas in
Fig.~\ref{figure4}(c)   shows   the   thermoelectric   figure-of-merit
$ZT\equiv  L_{PQ} L_{QP}/{\rm  det}  \bs{L}$ \cite{Thermoelectricity}.
One sees that while non-diagonal  elements $L_{QP}$ or $L_{PQ}$ can be
sometimes  negative, $L_{QQ}, L_{PP}$  are always  positive. Moreover,
$ZT$  reaches large  values  only for  disproportionate hopping  rates
$t_A$ and $t_B$.

On the other hand, systems interacting with environments inevitably
include irreversible processes, which breaking down time-reversibility
invariance and thus, the validity of the Onsager reciprocity relation
\cite{Alicki76}.  Fig.~\ref{figure4}(d) shows the $\gamma$ dependence
of $L_{PQ}/L_{QP}$, and we see that the relation is roughly linearly broken by
increasing $\gamma$, and thus, we conclude that the Onsager
reciprocity relation is satisfied only if there is a time reversible
dynamics for the total system, including the super-reservoirs, which
is the case in which the Onsager reciprocity relation is rigorously
proved~\cite{Philippe09}.  We recall that $\gamma\ll 1$ is one of the
necessary conditions for deriving the Lindblad equation, by taking a
partial trace of the unitary time evolution, thus $|L_{PQ}/L_{QP}-1|$
can be understood as an error-indicator due to the weak-coupling
assumption. We remark that the transport coefficients for the diatomic
chains (shown in Fig.~\ref{figure3}(d) and Fig.~\ref{figure4}) are
non-smooth functions of $t_B$ at the mono-atomic point $t_B=t_A$.

{\em Conclusion.-} We have introduced a model in the framework of open
quantum systems with finite meso-reservoirs which nicely
describes non-equilibrium steady states of quantum chains. We have
checked that this model has a regime which is robust with respect to
the strength of the Lindblad dissipators, and the occupation number
distributions of meso-reservoirs are close to the Fermi distribution, the difference being determined by the particle current.  
We
observed that the decay times of normal modes show a clear separation
into slow and fast decaying modes. In a certain regime we find a
possibility of strongly fluctuating currents (as a function of any
generic parameter) which is attributed to existence of well separated
intermediate (resonant) decay modes.  
We find
possibility of negative non-diagonal elements of the Onsager matrix
but confirm the positivity of the full Onsager matrix.  The Onsager
reciprocal relation was shown to be correct only for weak coupling
$\gamma\ll 1$, and the symmetry is broken linearly as a function of
$\gamma$.

The authors thank J. von Delft, D. Kosov, Y. Ohta, K. Saito and
M. \v{Z}nidari\v{c} for discussions on related subjects.  SA thanks
R. Soto for settings of computational environments and 
Fondecyt 3120254 for support.  TP acknowledges
supports by the grants P1-0044 and J1-2208 of the Slovenian Research
Agency.
TP and CM-M acknowledge partial support from Finlombarda project
``THERMOPOWER''.
FB and TP thanks international collaboration project Fondecyt 1110144.
Finally FB and SA thanks anillo ACT 127.
\bibliography{aps}

\end{document}